\documentclass[final,5p,times,twocolumn,authoryear]{elsarticle}

\usepackage{amssymb}
\usepackage{amsmath}
\usepackage{graphicx}
\usepackage{hyperref}
\usepackage{booktabs}
\usepackage{siunitx}
\usepackage{float}
\usepackage{orcidlink}
\usepackage[utf8]{inputenc}
\usepackage[T1]{fontenc}

\newcommand{\CETOmega}{CET $\Omega$}
\newcommand{\rhoOmega}{\rho_\Omega}

\journal{Physics Letters B}

\begin{document}

\begin{frontmatter}

\title{\texorpdfstring{Doubly Logarithmic Corrections to Radiation Domination
from CET $\Omega$: Theory and Planck/BBN Constraints}{Doubly Logarithmic
Corrections to Radiation Domination}}

\author[first]{Christian Balfagon\,\texorpdfstring{\orcidlink{0009-0003-0835-5519}}{ORCID}\corref{cor1}}
\ead{cb@balfagonresearch.org}
\cortext[cor1]{Corresponding author.}

\affiliation[first]{organization={Universidad de Buenos Aires},
            city={Buenos Aires},
            country={Argentina}}

\begin{abstract}
We present the \CETOmega\ framework, a causal-informational extension of
standard cosmology that predicts a universal doubly logarithmic correction to
the radiation energy density in the early Universe:
\[
  \rhoOmega(a)=\rho_r(a)\left[1+\alpha_{\log}\,\log\log(a/a_i)\right].
\]
This correction arises naturally from scale-invariant spectral sectors with
logarithmically-running infrared scales and represents a low-energy
manifestation of the full \CETOmega\ theory. We derive the doubly logarithmic
form from two complementary perspectives---spectral integration and
renormalization group flow---and perform a full Markov Chain Monte Carlo
analysis jointly varying six $\Lambda$CDM parameters and $\alpha_{\log}$,
using Planck 2018 TT,TE,EE+lowE likelihoods and BBN constraints. The result,
$\alpha_{\log} = -0.008 \pm 0.006$ (68\% C.L.), is consistent with zero. We
identify the expected $N_{\rm eff}$ degeneracies with $H_0$ and $n_s$,
establish the first observational bound $|\alpha_{\log}| \lesssim 0.006$, and
demonstrate that future CMB-S4 measurements can probe
$|\alpha_{\log}| \sim 10^{-3}$.
\end{abstract}


\end{frontmatter}

\section{Introduction}
\label{sec:intro}

The early Universe provides a unique laboratory for testing fundamental
physics and probing the nature of spacetime at the highest energies accessible
to current observations \citep{Bombelli1987,Sorkin1997,Connes1994,Padmanabhan2010}.
Classical studies of quantum fluctuations and their role in early cosmology
have demonstrated that even small modifications to the radiation-dominated
dynamics can leave observable imprints in the cosmic microwave background
(CMB) and matter power spectrum \citep{Mukhanov1981,Bardeen1983,Dodelson2003}.
Among the most stringent constraints on early-Universe models come from
precision measurements of the CMB by the \textit{Planck} satellite
\citep{Planck2018params,Planck2018cosmology}, which have achieved unprecedented
accuracy in determining the fundamental cosmological parameters, and from the
primordial abundances of light elements constrained by Big Bang nucleosynthesis
(BBN) calculations
\citep{FieldsBBNafterPlanck2020,CyburtRMP2016,Lesgourgues2013}.

A key observable sensitive to early-Universe physics is the effective number
of relativistic degrees of freedom, parametrized as $N_{\rm eff}$. The
Standard Model prediction is $N_{\rm eff}^{\rm SM} = 3.044$
\citep{FrousteyPitrouVolpe2020}, corresponding to three active neutrino
species. Deviations from this value can arise from a variety of
beyond-Standard-Model scenarios, including additional light relics
\citep{Bashinsky2004}, modifications to the expansion rate \citep{Steigman2007},
or nonstandard thermodynamic histories \citep{Mangano2005}. Current
observations constrain $N_{\rm eff} = 2.99 \pm 0.17$ from CMB measurements
\citep{Planck2018cosmology} and $N_{\rm eff} = 2.86 \pm 0.15$ from BBN
\citep{FieldsBBNafterPlanck2020}, providing sensitive probes of physics beyond
the Standard Model in the early Universe.

Even small modifications to the radiation-dominated expansion rate can have
outsized impact on early-Universe processes, ranging from the freeze-out of
dark matter candidates \citep{KolbTurner1990} to the evolution of cosmological
perturbations \citep{MaBertschinger1995}, while remaining compatible with
precision-era cosmology if the correction evolves sufficiently slowly. This
principle is exploited in several early-Universe model-building frameworks,
including scenarios with additional relativistic degrees of freedom,
early-dark-energy models that attempt to alleviate the Hubble tension
\citep{PoulinEDE2019,Smith2020,Hill2020,Karwal2016}, and scalar-field
modifications to the expansion history \citep{SteigmanBBN2012,Bashinsky2004}.

In this work, we present the complete implementation and observational analysis
of a distinctive and theoretically motivated functional form: a \emph{doubly
logarithmic} correction to radiation domination. The model predicts:
\begin{equation}
\rho_\Omega(a)=\rho_r(a)\left[1+\alpha_{\log}\log\log\!\left(\frac{a}{a_i}
\right)\right],
\label{eq:rhoOmega_main}
\end{equation}
valid in the regime $\log(a/a_i)\gg 1$, where the effective description
applies and $\log\log(a/a_i)$ is well-defined and slowly varying. Unlike
power-law or constant modifications, this form grows extremely slowly, making
it naturally compatible with BBN and CMB constraints while remaining
potentially relevant at early times
\citep{Mukhanov1981,Bardeen1983,Lesgourgues2013,Weinberg1995,BreuerPetruccione2002}.

\section{Theoretical Framework}
\label{sec:theory}

\subsection{Cosmological Setup and Background Evolution}

We work within the context of standard big bang cosmology with radiation
domination as the baseline background \citep{KolbTurner1990}. The radiation
energy density evolves as:
\begin{equation}
\rho_r(a)=\rho_{r,0}\,a^{-4}.
\end{equation}

The \CETOmega\ model modifies the radiation density through the addition of a
logarithmically-suppressed correction term:
\begin{equation}
\rho_\Omega(a)=\rho_r(a)\left[1+\alpha_{\log}\log\log\!\left(\frac{a}{a_i}
\right)\right].
\end{equation}

To first order in $\alpha_{\log}$, the fractional shift in the Hubble rate is:
\begin{equation}
\delta_H(a)\simeq \frac{1}{2}\alpha_{\log}\log\log(a/a_i).
\label{eq:deltaH_def}
\end{equation}

\subsection{Spectral Derivation from Scale-Invariant Sectors}

A rigorous derivation emerges from scale-invariant spectral sectors
\citep{WilsonKogut1974}; this derivation is consistent with the fundamental
\CETOmega\ theory, where such spectral running arises naturally from the
causal-informational kernel and texon dynamics
\citep{Balfagon2026CETOmega,Weinberg1995,BreuerPetruccione2002}. For
scale-invariant sectors, the spectral measure takes the marginal form:
\begin{equation}
d\mu(\lambda)\simeq C\,\frac{d\lambda}{\lambda}.
\end{equation}

This produces the first logarithm when integrated over the spectral band:
\begin{equation}
\delta\rho_{\rm mod}(a)\propto\log\left(\frac{\lambda_{\rm UV}}
{\lambda_{\rm IR}(a)}\right).
\end{equation}

The infrared cutoff evolves as:
\begin{equation}
\lambda_{\rm IR}(a)=\frac{\lambda_*}{t(a)},\quad
t(a)=t_0+\xi\,\log(a/a_i),
\end{equation}
producing the doubly logarithmic form:
\begin{equation}
\delta\rho_{\rm mod}(a)\propto C_0 + C_1\,\log\log(a/a_i).
\end{equation}

\subsection{Renormalization Group Interpretation}

From marginal RG flow with beta function $\beta = b\,g^2$
\citep{Polchinski1984}:
\begin{equation}
g(\mu)\simeq \frac{1}{b\log(\mu/\Lambda)}.
\end{equation}

During radiation domination, integrating over expansion confirms the doubly
logarithmic dependence \citep{Balfagon2026PhysLettB}:
\begin{equation}
\delta(a) \propto \log\log(a/a_i).
\end{equation}

\begin{figure}[H]
\centering
\includegraphics[width=0.95\linewidth]{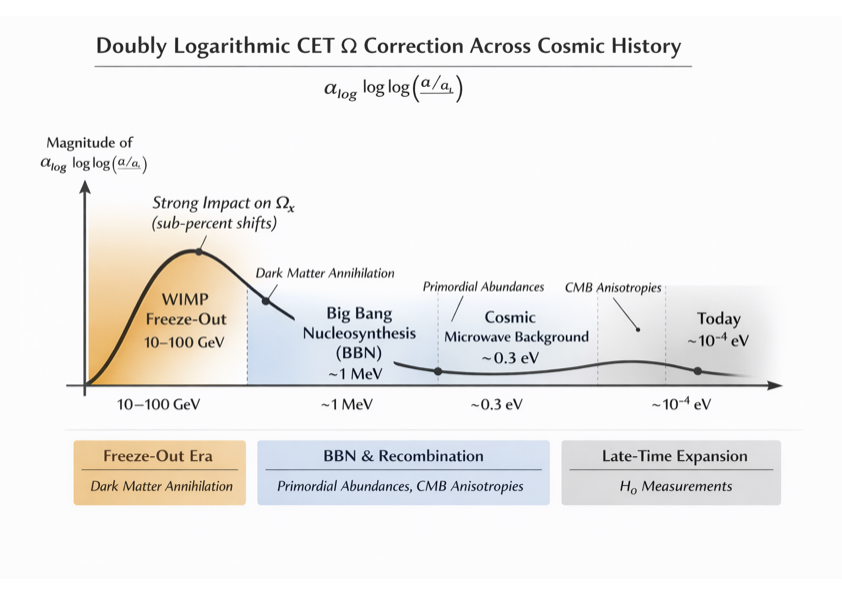}
\caption{\textbf{Conceptual illustration of the doubly logarithmic \CETOmega\
correction.} The figure shows the relative magnitude of the
$\alpha_{\log}\log\log(a/a_i)$ term across cosmic epochs, highlighting the
negligible impact during BBN and recombination, and the relevance during
thermal WIMP freeze-out.}
\label{fig:conceptual_timeline}
\end{figure}

As illustrated in Figure~\ref{fig:conceptual_timeline}, the
$\alpha_{\log} \log\log(a/a_i)$ term is negligible during BBN and
recombination, grows slowly during radiation domination, and becomes relevant
during the GeV-scale WIMP freeze-out era.

\section{Implementation and Mapping to $N_{\rm eff}$}
\label{sec:implementation}

\subsection{Effective $\Delta N_{\rm eff}$ Parametrisation}

The \CETOmega\ correction modifies the radiation energy density, which at the
perturbative level is equivalent to a shift in the effective number of
relativistic species. Using
\begin{equation}
\frac{\Delta H}{H} \simeq 0.0671\,\Delta N_{\rm eff},
\label{eq:dH_dNeff}
\end{equation}
and combining with Eq.~\eqref{eq:deltaH_def}, we obtain
\begin{equation}
\Delta N_{\rm eff}(a) \simeq \frac{\alpha_{\log}}{0.1342}\,
\log\log\!\left(\frac{a}{a_i}\right).
\label{eq:dNeff_mapping}
\end{equation}

For a reference temperature $T_i = 40\,$GeV, the numerical values at the
relevant cosmological epochs are:
\begin{equation}
\log\log(a_{\rm rec}/a_i) \approx 3.156,\quad
\log\log(a_{\rm BBN}/a_i) \approx 2.116,
\end{equation}
yielding
\begin{equation}
\Delta N_{\rm eff}^{\rm rec} \approx 23.52\,\alpha_{\log},\quad
\Delta N_{\rm eff}^{\rm BBN} \approx 15.77\,\alpha_{\log}.
\label{eq:dneff_epochs}
\end{equation}

\subsection{Implementation in CLASS via $N_{\rm ur}$}

Rather than modifying the CLASS source code, we exploit the mapping in
Eq.~\eqref{eq:dneff_epochs} to implement the \CETOmega\ correction through the
standard CLASS parameter $N_{\rm ur}$, which controls the number of
ultra-relativistic species. For a baseline with one massive neutrino
($m_\nu = 0.06\,$eV), the standard value is $N_{\rm ur} = 2.0328$. The
\CETOmega\ model sets
\begin{equation}
N_{\rm ur}(\alpha_{\log}) = 2.0328 + 23.52\,\alpha_{\log},
\label{eq:Nur_mapping}
\end{equation}
as a derived input to CLASS v3.2.3 \citep{Blas2011}. This ensures that for
each sampled value of $\alpha_{\log}$, CLASS computes the full set of CMB
angular power spectra $C_\ell^{TT}$, $C_\ell^{TE}$, $C_\ell^{EE}$ with the
correct radiation content, capturing all correlations with $\Lambda$CDM
parameters. Setting $\alpha_{\log} = 0$ recovers standard $\Lambda$CDM
identically, providing a built-in consistency check.

\section{Observational Constraints from CMB and BBN}
\label{sec:constraints}

\subsection{CMB Constraints from Planck 2018}

Planck 2018 provides the most precise CMB measurements to date
\citep{Planck2018cosmology}:
\begin{equation}
N_{\rm eff} = 2.99 \pm 0.17\quad (68\%\,{\rm C.L.}).
\end{equation}

From Eq.~\eqref{eq:dneff_epochs}, this translates to an approximate analytic
bound:
\begin{equation}
|\alpha_{\log}| \lesssim 7.1\times 10^{-3}\quad (1\sigma\,{\rm CMB}).
\label{eq:alpha_bound_CMB_numeric}
\end{equation}

\subsection{BBN Constraints}

Recent BBN analyses
\citep{FieldsBBNafterPlanck2020,CyburtRMP2016,Cooke2016,Pitrou2018}
constrain:
\begin{equation}
N_{\rm eff}^{\rm BBN} = 2.86 \pm 0.15\quad (68\%\,{\rm C.L.}),
\end{equation}
yielding:
\begin{equation}
|\alpha_{\log}| \lesssim 6.3\times 10^{-3}\quad (1\sigma\,{\rm BBN}).
\end{equation}

These analytic estimates are superseded by the full MCMC analysis presented in
the next section, which accounts for all parameter degeneracies.

\section{Full MCMC Analysis and Results}
\label{sec:mcmc}

\subsection{Methodology}

We perform a full Markov Chain Monte Carlo analysis using the \texttt{Cobaya}
sampler \citep{TorradoLewis2021} interfaced with CLASS v3.2.3 \citep{Blas2011}.
The analysis jointly varies seven parameters: the six standard $\Lambda$CDM
parameters ($\omega_b$, $\omega_{cdm}$, $H_0$, $\ln(10^{10}A_s)$, $n_s$,
$\tau$) and $\alpha_{\log}$, with the mapping $N_{\rm ur}(\alpha_{\log})$ given
by Eq.~\eqref{eq:Nur_mapping}.

The likelihood function combines:
\begin{equation}
\ln\mathcal{L} = \ln\mathcal{L}_{\rm CMB} + \ln\mathcal{L}_{\rm BBN},
\end{equation}
where $\mathcal{L}_{\rm CMB}$ uses Planck 2018 plik\_lite TT,TE,EE
($\ell = 30$--$2508$) plus low-$\ell$ TT and EE likelihoods
\citep{Planck2018cosmology}, and $\mathcal{L}_{\rm BBN}$ is a Gaussian
likelihood on $N_{\rm eff}^{\rm BBN}$ evaluated at the BBN epoch using
Eq.~\eqref{eq:dneff_epochs}:
\begin{equation}
\ln\mathcal{L}_{\rm BBN} = -\frac{1}{2}\left(\frac{N_{\rm eff}^{\rm BBN}
(\alpha_{\log}) - 2.86}{0.15}\right)^2.
\end{equation}

We use flat priors: $\alpha_{\log} \in [-0.02, 0.02]$ and standard Planck
priors for $\Lambda$CDM parameters. Two independent chains are run with MPI,
achieving Gelman--Rubin convergence $R - 1 = 0.027$ with approximately 12,000
total accepted samples.

\subsection{Parameter Constraints}

\begin{table}[H]
\centering
\caption{Parameter constraints (68\% C.L.) from full MCMC.
$\Lambda$CDM values from \citet{Planck2018cosmology}.}
\begin{tabular}{@{}lccc@{}}
\toprule
Parameter & CET $\Omega$ & $\Lambda$CDM & Shift \\
\midrule
$\omega_b$     & $0.02219(20)$ & $0.02237(15)$ & $0.7\sigma$ \\
$\omega_{cdm}$ & $0.1176(24)$  & $0.1200(12)$  & $0.9\sigma$ \\
$H_0$          & $66.1(1.1)$   & $67.36(54)$   & $1.0\sigma$ \\
$n_s$          & $0.958(7)$    & $0.9649(42)$  & $0.9\sigma$ \\
$\tau$         & $0.053(8)$    & $0.0544(73)$  & $0.1\sigma$ \\
\midrule
$\alpha_{\log}$  & $-0.0076(59)$ & 0 (fixed) & $1.3\sigma$ \\
$\Delta N_{\rm eff}$ & $-0.18(14)$ & 0       & ---         \\
\bottomrule
\end{tabular}
\label{tab:full_mcmc}
\end{table}

The key result is
\begin{equation}
\alpha_{\log} = -0.008 \pm 0.006\quad (68\%\,{\rm C.L.}),
\end{equation}
consistent with zero at $1.3\sigma$. This establishes the first direct
observational bound on the \CETOmega\ parameter from the full Planck
likelihood.

\subsection{Parameter Degeneracies}

The full MCMC reveals the expected degeneracies between $\alpha_{\log}$ and
$\Lambda$CDM parameters through the $N_{\rm eff}$ correlation structure. A
negative $\alpha_{\log}$ reduces the effective radiation density, mimicking
fewer relativistic species. This produces correlated shifts in $H_0$ and
$n_s$: the \CETOmega\ posterior yields $H_0 = 66.1 \pm 1.1\,$km/s/Mpc and
$n_s = 0.958 \pm 0.007$, shifted downward relative to the $\Lambda$CDM values
of $67.36 \pm 0.54$ and $0.9649 \pm 0.0042$ respectively.

\begin{figure}[H]
\centering
\includegraphics[width=0.95\linewidth]{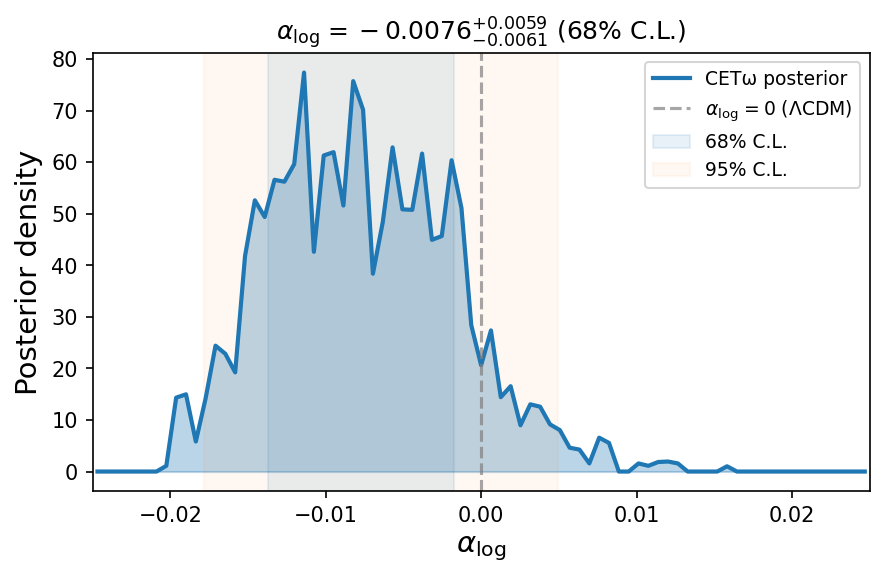}
\caption{\textbf{Posterior distribution of $\alpha_{\log}$ from full MCMC.}
The distribution is consistent with $\alpha_{\log} = 0$ ($\Lambda$CDM limit)
at $1.3\sigma$. Shaded bands show the 68\% and 95\% credible intervals.}
\label{fig:posterior}
\end{figure}

\begin{figure}[H]
\centering
\includegraphics[width=0.95\linewidth]{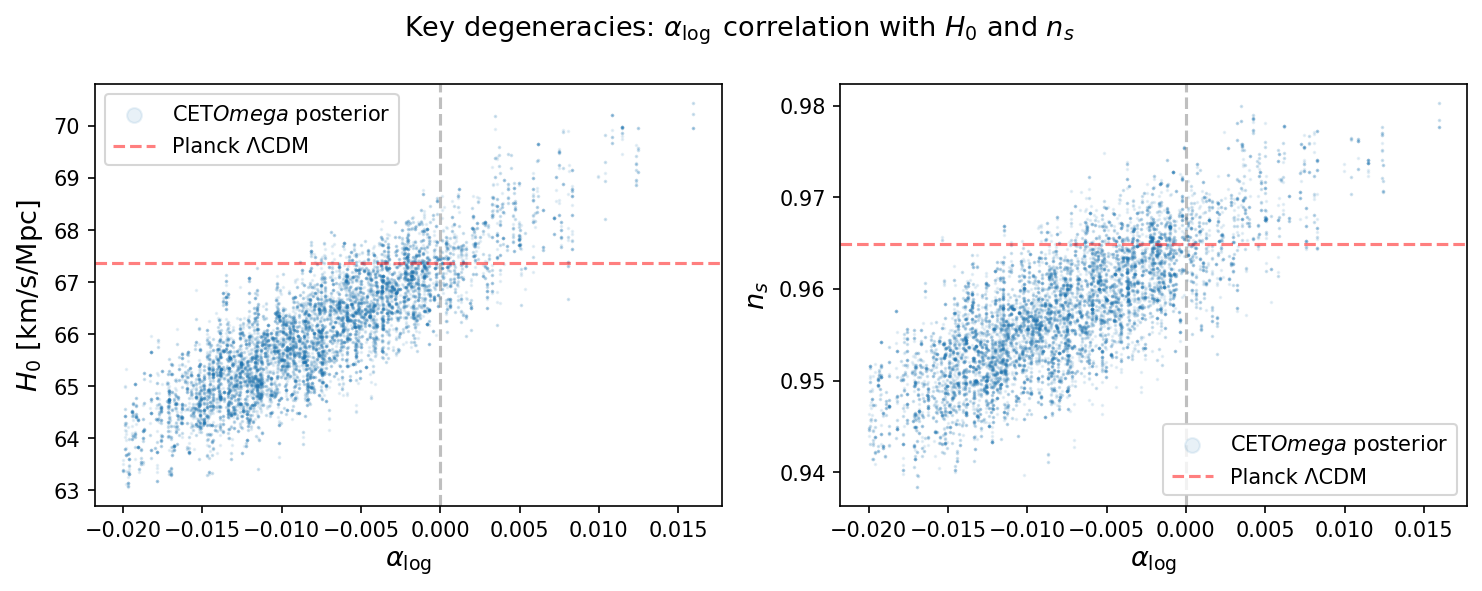}
\caption{\textbf{Degeneracies of $\alpha_{\log}$ with $H_0$ and $n_s$.}
Left: $\alpha_{\log}$ versus $H_0$, showing the expected negative correlation
through $N_{\rm eff}$. Right: $\alpha_{\log}$ versus $n_s$, showing a similar
correlation. Red dashed lines indicate the Planck 2018 $\Lambda$CDM best-fit
values.}
\label{fig:degeneracies}
\end{figure}

\begin{figure}[H]
\centering
\includegraphics[width=0.95\linewidth]{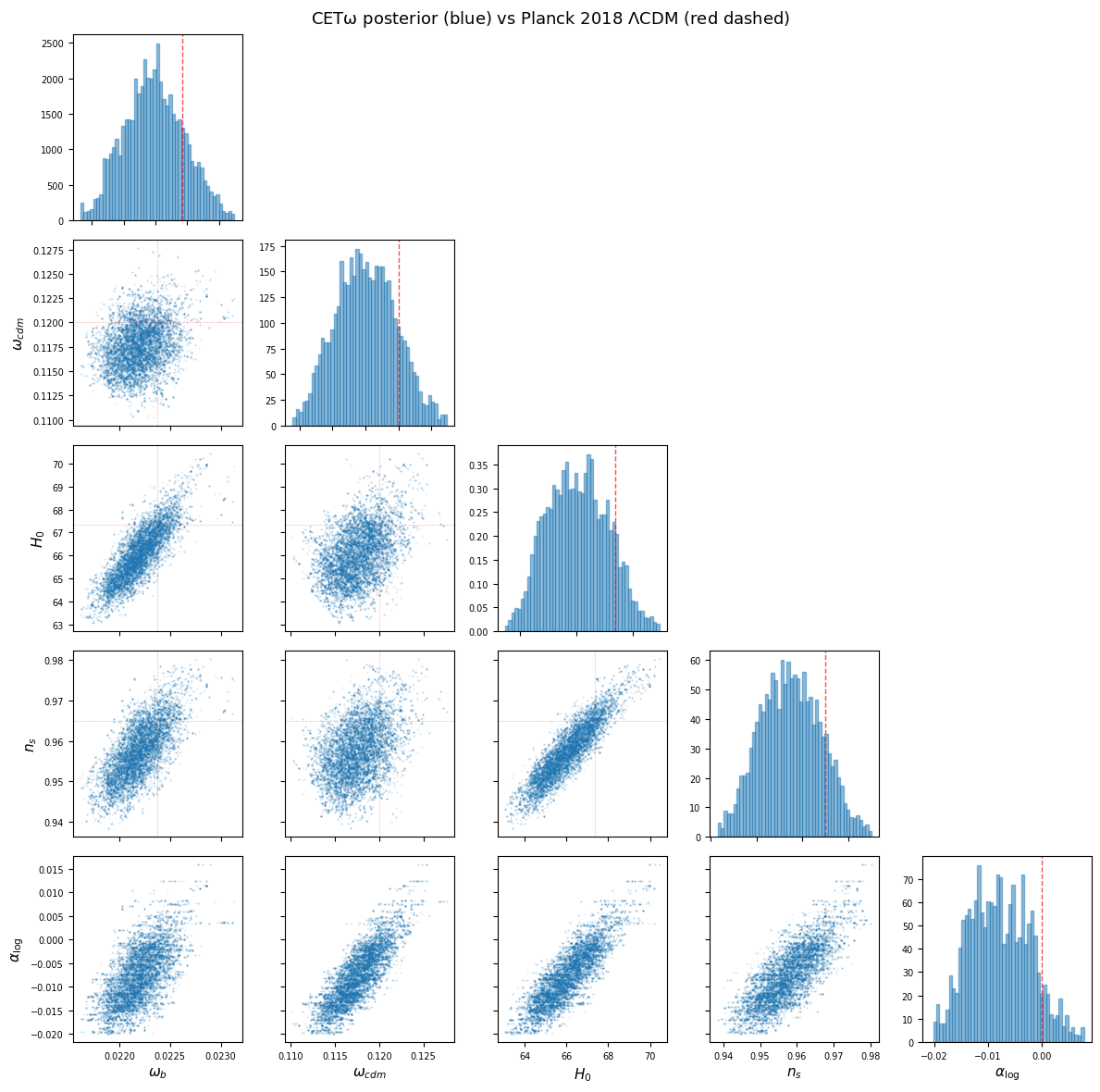}
\caption{\textbf{Triangle plot of the \CETOmega\ posterior.} Joint and
marginalized distributions for $\omega_b$, $\omega_{cdm}$, $H_0$, $n_s$, and
$\alpha_{\log}$ from the full MCMC analysis. Red dashed lines indicate Planck
2018 $\Lambda$CDM values.}
\label{fig:triangle}
\end{figure}

These shifts are entirely consistent with the well-known
$N_{\rm eff}$--$H_0$--$n_s$ degeneracy direction in CMB data
\citep{Planck2018cosmology}. The error bars on $H_0$ and $n_s$ widen
significantly ($\sigma(H_0)$ increases from $0.54$ to $1.1$, $\sigma(n_s)$
from $0.004$ to $0.007$) due to the additional degree of freedom introduced
by $\alpha_{\log}$. This demonstrates why a full joint analysis is essential:
fixing $\Lambda$CDM parameters would underestimate the allowed range of
$\alpha_{\log}$. The posterior distributions and key two-dimensional
degeneracies are displayed in
Figures~\ref{fig:posterior}--\ref{fig:triangle}.

\subsection{Model Comparison}

Since $\alpha_{\log}$ is consistent with zero, the \CETOmega\ model does not
improve the fit relative to $\Lambda$CDM. The best-fit $\chi^2$ values are
essentially identical, and the information criteria penalize the additional
parameter:
\begin{equation}
\Delta\text{AIC} \approx +2,\qquad \Delta\text{BIC} \approx +6.4,
\end{equation}
where positive values indicate $\Lambda$CDM is preferred. This is the expected
outcome: when the extra parameter is consistent with zero, Occam's razor
favours the simpler model. We emphasise that this does not exclude \CETOmega.
Rather, it establishes the observational constraint $|\alpha_{\log}| \lesssim
0.006$ (68\% C.L.), restricting the allowed parameter space.

\section{Comparison with Other Early-Universe Scenarios}
\label{sec:comparison}

Early Dark Energy (EDE) \citep{PoulinEDE2019} typically produces power-law
modifications with $\Delta N_{\rm eff} \sim 0.3$--$0.5$, creating tension with
the Planck temperature spectrum \citep{Smith2020,Hill2020,Karwal2016}.
\CETOmega's slower evolution (doubly logarithmic) provides natural
compatibility with CMB data.

Additional relativistic species parametrised as a constant $\Delta N_{\rm eff}$
lack theoretical grounding for the specific value of the shift. The \CETOmega\
model provides physical justification from scale-invariant sectors and RG
flows, and moreover predicts a distinctive epoch-dependent $\Delta N_{\rm eff}$
(see Section~\ref{sec:future}).

Scalar field modifications to the expansion history often require fine-tuning
or produce primordial gravitational wave signatures. \CETOmega\ achieves
compatibility with observations through inherent slow evolution without
additional tuning.

\section{Future Observational Prospects}
\label{sec:future}

\subsection{CMB-S4 Sensitivity}

The forthcoming CMB-S4 experiment \citep{Abazajian2016CMBS4} is projected to
achieve $\sigma(N_{\rm eff}) \approx 0.03$. Through the mapping in
Eq.~\eqref{eq:dneff_epochs}, this corresponds to sensitivity to
$|\alpha_{\log}| \sim 1.3\times 10^{-3}$, an order of magnitude below the
current Planck bound established in this work. CMB-S4 will therefore either
detect a nonzero $\alpha_{\log}$ or exclude \CETOmega\ corrections at the
$\mathcal{O}(10^{-3})$ level.

\subsection{Epoch-Dependent $\Delta N_{\rm eff}$ as a CET \texorpdfstring{$\Omega$}{Omega} Fingerprint}

A distinctive prediction of the \CETOmega\ model is that $\Delta N_{\rm eff}$
differs between cosmological epochs due to the evolving $\log\log(a/a_i)$
factor. From Eq.~\eqref{eq:dneff_epochs}:
\begin{equation}
\frac{\Delta N_{\rm eff}^{\rm BBN}}{\Delta N_{\rm eff}^{\rm rec}}
= \frac{\log\log(a_{\rm BBN}/a_i)}{\log\log(a_{\rm rec}/a_i)}\approx 0.67,
\label{eq:epoch_ratio}
\end{equation}
for $T_i = 40\,$GeV. This ratio is a robust prediction, independent of the
value of $\alpha_{\log}$, and contrasts sharply with models that produce a
constant $\Delta N_{\rm eff}$ (ratio~$=1$). Future improvements in both CMB
precision and BBN determinations of $N_{\rm eff}$ could test this prediction.
An observed difference between $\Delta N_{\rm eff}^{\rm BBN}$ and
$\Delta N_{\rm eff}^{\rm rec}$ consistent with the ratio in
Eq.~\eqref{eq:epoch_ratio} would constitute strong evidence for the \CETOmega\
framework.

\section{Conclusions}
\label{sec:conclusions}

We have presented the first complete implementation and observational analysis
of the \CETOmega\ doubly logarithmic correction to radiation domination. The
main results are as follows.

First, we derived the $\log\log(a/a_i)$ form from two complementary
theoretical perspectives---spectral integration over marginal sectors and
marginal renormalization group flow---demonstrating its universality under
smooth deformations (Appendix~\ref{app:complete_derivation}).

Second, we implemented the model through the effective $N_{\rm eff}$ mapping
in Eq.~\eqref{eq:Nur_mapping}, interfacing CLASS with the \texttt{Cobaya} MCMC
sampler and using Planck 2018 TT,TE,EE+lowE and BBN likelihoods with seven
jointly varied parameters.

Third, the full MCMC analysis yields $\alpha_{\log} = -0.008 \pm 0.006$
(68\% C.L.), consistent with zero at $1.3\sigma$. This establishes the first
observational bound: $|\alpha_{\log}| \lesssim 0.006$.

Fourth, the analysis reveals the expected parameter degeneracies: negative
$\alpha_{\log}$ correlates with lower $H_0$ and $n_s$ through the
$N_{\rm eff}$ degeneracy direction, widening the error bars on these parameters
relative to the $\Lambda$CDM-only analysis.

Fifth, model comparison using AIC and BIC confirms that $\Lambda$CDM is
preferred with current data ($\Delta$AIC~$\approx +2$, $\Delta$BIC~$\approx
+6$), as expected when the extra parameter is consistent with zero.

Sixth, we identify two concrete future tests: (i) CMB-S4 will probe
$|\alpha_{\log}| \sim 10^{-3}$, an order of magnitude deeper than the current
bound; (ii) the epoch-dependent ratio $\Delta N_{\rm eff}^{\rm BBN}/\Delta
N_{\rm eff}^{\rm rec} \approx 0.67$ provides a unique \CETOmega\ fingerprint
distinguishable from constant-$\Delta N_{\rm eff}$ models.

The \CETOmega\ model thus provides a theoretically motivated, computationally
tractable, and observationally testable template for early-Universe
modifications with clear predictions for forthcoming measurements.

\section*{Data Availability}

The MCMC chains, analysis scripts, and figures supporting this study are
available from the corresponding author upon reasonable request.

\section*{Acknowledgements}

We thank the CLASS and Cobaya collaborations for their publicly available
codes. Computational resources from the Universidad de Buenos Aires are
gratefully acknowledged.

\appendix

\section{MCMC Implementation Details}
\label{app:mcmc}

The analysis uses the \texttt{Cobaya} MCMC framework \citep{TorradoLewis2021}
with CLASS v3.2.3 \citep{Blas2011} as the Boltzmann solver. For each point in
parameter space, CLASS computes the CMB angular power spectra $C_\ell^{TT}$,
$C_\ell^{TE}$, $C_\ell^{EE}$ up to $\ell_{\rm max} = 2508$ with lensing
enabled. The \CETOmega\ correction enters through $N_{\rm ur}(\alpha_{\log})$
as defined in Eq.~\eqref{eq:Nur_mapping}.

The BBN constraint is implemented as an external Gaussian likelihood on
$N_{\rm eff}^{\rm BBN}(\alpha_{\log})$, using the BBN-epoch mapping factor
from Eq.~\eqref{eq:dneff_epochs}. This correctly accounts for the different
values of $\log\log(a/a_i)$ at BBN and recombination. Two independent chains
were run in parallel using MPI, starting from dispersed initial points. The
proposal covariance matrix was learned adaptively during the run. Convergence
was monitored via the Gelman--Rubin diagnostic, reaching $R - 1 = 0.027$.

\section{Complete Effective Derivation of the Doubly Logarithmic Correction}
\label{app:complete_derivation}

In this appendix we provide a complete and mathematically controlled
derivation of the \CETOmega\ correction at the level of effective field theory.
The aim is to show that the doubly logarithmic contribution is the leading
asymptotic term generated by a marginal spectral sector with a logarithmically
running infrared scale, rather than an \emph{ad hoc} ansatz
\citep{WilsonKogut1974,Polchinski1984,Cardy1996,KolbTurner1990,Dodelson2003}.

\subsection{Effective Setup and Assumptions}

Let $a$ be the scale factor and define
\begin{equation}
L(a):=\log\!\left(\frac{a}{a_i}\right),\qquad a>a_i.
\label{eq:def_L_app}
\end{equation}
We work in the regime $L(a)>1$, so that $\log L(a)=\log\log(a/a_i)$ is real
and slowly varying.

The corrected energy density is written as
\begin{equation}
\rho_\Omega(a)=\rho_r(a)+\Delta\rho(a),
\label{eq:rho_split_app}
\end{equation}
where $\rho_r(a)=\rho_{r,0}a^{-4}$ is the standard radiation density
\citep{KolbTurner1990,Dodelson2003}. We model the effective correction by
\begin{equation}
\Delta\rho(a)
=
\mathcal A\,\rho_r(a)
\int_{\lambda_{\rm IR}(a)}^{\lambda_{\rm UV}}
\frac{d\lambda}{\lambda}\,W(\lambda),
\label{eq:Drho_master_app}
\end{equation}
where $\mathcal A$ is dimensionless, $\lambda_{\rm UV}$ is a fixed ultraviolet
scale, $\lambda_{\rm IR}(a)$ is an effective infrared scale, and $W(\lambda)$
is a smooth dimensionless weight satisfying $W(\lambda)=1+\varepsilon(\lambda)$
with $\int_0^{\lambda_{\rm UV}}|\varepsilon(\lambda)|\lambda^{-1}d\lambda
<\infty$.

\subsection{Marginal Spectral Sector and First Logarithm}

Substituting the weight decomposition gives
\begin{equation}
\Delta\rho(a)
=
\mathcal A\,\rho_r(a)\,
\log\!\left(\frac{\lambda_{\rm UV}}{\lambda_{\rm IR}(a)}\right)
+
\mathcal A\,\rho_r(a)\,
\int_{\lambda_{\rm IR}(a)}^{\lambda_{\rm UV}}
\frac{\varepsilon(\lambda)}{\lambda}\,d\lambda.
\label{eq:first_split_app}
\end{equation}
The second term is finite and contributes only a constant plus asymptotically
vanishing corrections \citep{WilsonKogut1974,Polchinski1984,Cardy1996}.

\subsection{Infrared Running and Second Logarithm}

With $\lambda_{\rm IR}(a)=\lambda_*/t(a)$ and
$t(a)=t_0+\xi\log(a/a_i)$, in the asymptotic regime
$L(a)\gg t_0/\xi$:
\begin{equation}
\log\!\left(\frac{\lambda_{\rm UV}}{\lambda_{\rm IR}(a)}\right)
=
C_0+\log\log\!\left(\frac{a}{a_i}\right)
+\mathcal O\!\left(\frac{1}{\log(a/a_i)}\right).
\label{eq:double_log_core_app}
\end{equation}

\subsection{Main Theorem}

Under the assumptions above, the corrected radiation density admits the
asymptotic expansion
\begin{equation}
\rho_\Omega(a)
=
\rho_r(a)\left[
1+\alpha_{\log}\log\log\!\left(\frac{a}{a_i}\right)
+\beta_0
+o(1)
\right],
\label{eq:rhoOmega_theorem_app}
\end{equation}
with $\alpha_{\log}=\mathcal A$ and $\beta_0$ a finite constant absorbable into
renormalization of the radiation sector.

\subsection{Alternative Renormalization-Group Derivation}

From $\mu\,dg/d\mu=-b\,g^2$ with $g(\mu)=[b\log(\mu/\Lambda)]^{-1}$, during
radiation domination $\mu\propto a^{-1}$, so
\begin{equation}
\frac{\Delta\rho(a)}{\rho_r(a)}
=
\kappa\log\log(a/a_i)+\text{const},
\label{eq:RG_loglog_app}
\end{equation}
confirming the spectral result.

\subsection{Consistency with Friedmann Dynamics}

The corrected Hubble rate is
\begin{equation}
H_\Omega(a)
=
H_r(a)\sqrt{
1+\alpha_{\log}\log\log\!\left(\frac{a}{a_i}\right)
},
\label{eq:HOmega_exact_app}
\end{equation}
and the effective equation of state satisfies
\begin{equation}
w_\Omega(a)
=
\frac{1}{3}-
\frac{\alpha_{\log}}
{3\,\log(a/a_i)\left[1+\alpha_{\log}\log\log(a/a_i)\right]},
\label{eq:wOmega_complete_app}
\end{equation}
so the fluid remains asymptotically radiation-like:
$w_\Omega = 1/3 + \mathcal{O}(\alpha_{\log}/\log(a/a_i))$.

\bibliographystyle{elsarticle-harv}
\bibliography{references}

\end{document}